\documentstyle[epsfig,12pt]{article}
\input{epsf.sty}
\newcommand{\pp}{{_{I\hspace{-0.2em}P}}}
\begin{document}
\parskip 0.3cm
\begin{titlepage}
\begin{flushright}
CERN-TH/98-109\\
\end{flushright}

\begin{centering}
{\large \bf QCD Analysis of Diffractive and }\\
{\large \bf Leading-Proton DIS Structure Functions} \\
{\large \bf in the Framework of Fracture Functions}

 \vspace{.9cm}
{\bf  D. de Florian}
\vspace{.05in}
\\ {\it Theoretical Physics Division, CERN, CH 1211 Geneva 23, Switzerland}\\
e-mail: Daniel.de.Florian@cern.ch \\
 \vspace{.4cm}
{\bf R. Sassot} \\
\vspace{.05in}
{\it  Departamento de F\'{\i}sica, 
Universidad de Buenos Aires \\ 
Ciudad Universitaria, Pab.1 
(1428) Buenos Aires, Argentina} \\
\vspace{1.5cm}
{ \bf Abstract}
\\ 
\bigskip
\end{centering}
{\small We present a combined QCD analysis of recent data produced by the H1 and ZEUS collaborations on the diffractive and leading-proton deep inelastic positron--proton scattering structure functions, $F_2^{D(3)}$ and $F_2^{LP(3)}$, respectively. It is shown that the QCD framework for semi-inclusive processes,  based on fracture functions, allows a unified treatment of both diffractive and leading-proton processes, and offers a precise perturbative QCD description for them, alternative to those that rely on model-dependent assumptions, such as Regge factorization, or other non-perturbative approaches.}

\vspace{0.2in}
\vfill
\begin{flushleft}
CERN-TH/98-109\\
April 1998 \hfill
\end{flushleft}
\end{titlepage}
\vfill\eject

\noindent{\large \bf Introduction}\\

In recent years considerable attention has been paid to deep inelastic lepton--proton scattering processes with either an identified leading-proton in the final state \cite{ZLP,CART,H1LP,TAPP}, or of a diffractive nature, presumably dominated by the single
dissociative process $\gamma^*p \rightarrow Xp$ \cite{H1D,ZD}. The two kinds of events differ considerably in the kinematical regions where they are produced and also in the non-perturbative mechanisms or models used to describe them, such as meson exchange in the first case, and `pomeron+reggeon' exchange in the second. Consequently, their description is usually done within completely unrelated frameworks and in terms of different structure functions, the 
leading-proton and the diffractive deep inelastic scattering positron--proton structure functions, $F_2^{D(3)}$ and $F_2^{LP(3)}$, respectively.

However, if the experiment under consideration is able to identify the final-state proton in the diffractive processes, as in the ZEUS measurements, or  if there is confidence in the dominance of the single dissociative process $\gamma^*p \rightarrow Xp$, both kinds of events can be thought as of semi-inclusive nature, with identical final-state particles produced in the target fragmentation region. From this point of view, the perturbative QCD framework for their description must be identical, with only one factorized observable, even though the specific models for their non-perturbative features are completely different.

In perturbative QCD, the most appropriate description, for semi-inclusive DIS events in which the identified final-state hadron is produced in the target fragmentation region, is the one that includes fracture functions \cite{VEN}. These functions can  be  understood essentially as parton densities in an already fragmented target, and  there are different reasons for which they are necessary ingredients. 
In the first place, the more familiar description of semi-inclusive events, in terms of parton densities and fragmentation functions, refers essentially to only a limited set of current fragmentation processes; in leading order, final-state hadrons are   taken into account only if they are produced in the backward direction. Going to higher orders, a breakdown of hard factorization is found unless certain kinematical regions are subtracted from the cross section in an
arbitrary and process-dependent way, usually done by imposing cuts on the 
final-state variable $z$ \cite{API}.

Fracture functions solve all the above-mentioned problems, introducing naturally the leading-order description for target fragmentation events in the forward direction, and also a generalization to higher orders, allowing a consistent factorization of collinear divergences \cite{GRAU}. At the same time, the formalism provides
not only the motivation for establishing a connection between the kinematical regions of these kinds of processes, but also rigorous predictions about the perturbative QCD behaviour of the cross-sections. A priori this behaviour  is not the same as that of the ordinary structure functions due to appearence of an inhomogeneous term in the Altarelli--Parisi equations for fracture functions \cite{VEN,CGT}.

As for structure functions in totally inclusive deep inelastic scattering, QCD does not predict the shape of fracture functions unless it is known at a given initial scale. This non-perturbative information has to be obtained from the experiment, and, eventually, can be parametrized finding inspiration in non-perturbative models, as  for ordinary structure functions.
Up to now, the formalism of fracture functions has been successfully applied to describe forward neutron data from the ZEUS experiment \cite{NEUTR}, using a 
model as non-perturbative input.

In this paper we present results from a QCD global analysis of  recent semi-inclusive deep inelastic scattering data produced by H1  using the framework of fracture functions. It is shown that this approach unifies the description of   diffractive and leading-proton phenomena and allows a perturbative QCD description without the usual assumptions about approximate Regge factorization. 

As a result of this analysis, we also present a parametrization for the fracture function that characterizes the underlying semi-inclusive process at a given initial scale, as extracted from H1  data. The
scale dependence of these data is found to be in remarkable agreement with the
one predicted for the fracture function,  driven by the corresponding evolution equations. The parametrization is also used to compute other observables measured by ZEUS, not included in the fit, finding also an outstanding agreement with the data. 

In the following section we show the relation between leading-proton and diffractive structure functions, introducing the concept of fracture functions  in this context. Then, motivated by known non-perturbative models, we propose a parametric form for fracture functions at an initial scale, and fix their parameters by means of a leading order (LO) global QCD fit to H1 data. Finally we compute, with the resulting parametrization, semi-inclusive observables  measured by ZEUS and present our conclusions. \\

\noindent{\large \bf Definitions}\\

It is customary to define the leading-proton structure function $F_2^{LP (3)}$
from the corresponding triple-differential deep inelastic scattering cross section:
\begin{equation}
\frac{d^3 \sigma^{LP}}{dx\,dQ^2\,d\xi} \equiv \frac{4\pi \alpha^2}{x\,Q^4}\left( 1-y+\frac{y^2}{2}\right) F_2^{LP (3)}(x,Q^2,\xi) \, ,
\end{equation} 
with the usual
kinematical variables
\begin{equation}
Q^2=-q^2 \,\,\,\,\,\,\,\, x=\frac{Q^2}{2P\cdot q} \,\,\,\,\,\,\,\, \xi=\frac{q \cdot(P-P')}{q \cdot P}
\,\,\,\,\,\,\,\, \beta = \frac{Q^2}{2q \cdot (P-P')}=\frac{x}{\xi} \, . \end{equation}
Here $q$, $P$, and $P'$ are the momenta of the virtual boson, of the incident proton, and of the final state proton, respectively, and $y=Q^2/(x\, s)$. Both the cross section and the structure function defined in this way, include an implicit integration over a given range
for the small transverse momentum of the final-state proton. 

Even though the processes accounted for in eq. (1) are explicitly of a  semi-inclusive nature, the formulation based on the leading-proton structure
function is used instead of the usual approach for semi-inclusive deep inelastic  scattering 
\begin{equation}
\frac{d^3\sigma_{curr.}^p}{dx\,dQ^2\,dz} \simeq \frac{4\pi \alpha^2}{x Q^4}\left( 1-y+\frac{y^2}{2}\right) x\sum_i e^2_i\,q_i(x,Q^2) \times D^p_i (z,Q^2)  
\end{equation}
in terms of parton distributions and fragmentation functions, because the last one only takes into account hadrons produced in the current fragmentation region and thus not contributing to the forward leading hadron observables. 

These problems, and those related to factorization of collinear singularities at higher orders, are overcome, however, if the complete perturbative framework
for semi-inclusive processes is taken into account, for which the LO expression for the production of very forward hadrons is  \footnote{For the sake of simplicity we define the variable $z\equiv E_{P'}/E_{P}$ as the  ratio between the energies of the final-state proton and the proton beam in the centre of mass of the virtual photon--proton system. For very forward protons then, $1-\xi \simeq z $.}       
\begin{eqnarray}
\frac{d^3\sigma^p_{target}}{dx\,dQ^2\,dz} =\frac{4\pi \alpha^2}{x\,Q^4}\left( 1-y+\frac{y^2}{2}\right)   \sum_i e^2_i x M_i^{p/p}(x,z,Q^2)\, ,
 \end{eqnarray}
where $M_i^{p/p}(x,z,Q^2)$ is the fracture function that accounts for target fragmentation processes.

Fracture functions can be thought of in terms of the elements of any lowest-order picture for hadronization, for example as the product of a flux of exchanged `reggeons' times their structure functions, or more formally as an ingredient of the perturbative QCD treatment: the non-perturbative parton distributions of a proton fragmented into a proton. The latter choice can be generalized to higher orders, allowing a consistent analysis of the scale dependence of the semi-inclusive  cross section. This scale dependence is driven by inhomogeneous Altarelli--Parisi equations
\begin{eqnarray}
\frac{\partial M^h_i(x,z,Q^2)}{\partial \mbox{log}Q^2} =\frac{\alpha_s(Q^2)}{2\pi}\int^1_{x/(1-z)} \frac{dy}{y} P_{ij}(y) M^h_j\left( \frac{x}{y},z,Q^2\right) \\
+  \frac{\alpha_s(Q^2)}{2\pi}\int^{x/(x+z)}_x \frac{dy}{x(1-y)} \hat{P}_{ijl}(y)q_j \left(\frac{x}{y},Q^2\right) D^h_l\left( \frac{zy}{x(1-y)},Q^2\right) \, ,  \nonumber 
\end{eqnarray} 
where $ P_{ij}(y)$ and $\hat{P}_{ijl}(y)$ are the regularized and real   splitting functions,
reflecting the fact that the evolution is not only driven by the emission of collinear partons from those found in the target (homogeneous evolution), but also by
the fragmentation of partons radiated from the one struck by the virtual probe (inhomogeneous evolution).

Defining the equivalent to $F_2$ for fracture functions, i.e.
\begin{equation}
M_2^{p/p}(x,z,Q^2)\equiv x \sum_i e_i^2 M_i^{p/p}(x,z,Q^2),
\end{equation}
and taking into account the shift from $z$ to $\xi$, the relation between this function and the leading-proton structure function is quite apparent.
In ref. \cite{NEUTR}, the use of fracture functions for the description of leading-hadrons produced in the target fragmentation region has already been discussed in relation to the analysis of very forward neutrons observed by the ZEUS collaboration. There it was shown that, neglecting contributions beyond LO coming from the current fragmentation region, what is usually defined as the leading-neutron structure function $F_2^{LN (3)}$ is just the fracture function contribution to the semi-inclusive cross section. 

 Analogously to eq. (1), the differential cross section for diffractive deep inelastic scattering is usually written in terms of the diffractive structure function $F_2^{D(3)}$         
\begin{equation}
\frac{d^3 \sigma^{D}}{d\beta\,dQ^2\,dx_{\pp}} \equiv \frac{4\pi \alpha^2}{\beta\,Q^4}\left( 1-y+\frac{y^2}{2}\right) F_2^{D (3)}(\beta,Q^2,x_{\pp}) \, ,
\end{equation} 
where $x_{\pp}\equiv \xi$, and the variable $\beta$ is used instead of $x$.
 As in the previous case, an integration over the small transverse momentum of the final-state proton is implied. In this context it is customary to use the variable $t=(P-P')^2$ instead of the transverse momentum. When this last cross section is dominated by the single dissociative process $\gamma^*p \rightarrow Xp$, implying that there is a proton in the final state, the contributions to it are again given by the same fracture function in (4), even though in a completely different kinematical region\footnote{ The integration over the measured range of the variable $t$ can straightforwardly be circumvented  by defining `generalized' \cite{GRAZ} fracture functions with an explicit dependence on that variable and obeying homogeneous evolution equations, analogously to the diffractive structure function $F_2^{D(4)}$. However, in the present analysis we restrict ourselves to ordinary fracture functions.}. The diffractive region is given by small values of $x_{\pp}$ ($x_{\pp}<
0.1$),  whereas leading-proton data are associated with larger values of $x_{\pp}$ (
$x_{\pp}>0.1$). The available data also differ considerably in the ranges covered by the variable $\beta$, $0.04<\beta<0.9$ in the case of diffractive structure functions, and very small values ($\beta<0.03$) for leading-proton data.

Different kinematical 
regions correspond to different behaviours and also to different underlying 
models. The leading-proton structure function has been measured by the H1 collaboration \cite{H1LP,TAPP}, and has also been compared with predictions of different  mechanisms, such as meson exchange and soft colour interaction models for example, implemented in event generator programmes   
 \cite{ING,JUNG}, none of which reproduced integrally the main features of the data \cite{H1LP}. 

 The standard interpretation for the diffractive cross section is given in terms
of `pomeron' exchange. In this framework, different model estimates, and even QCD-inspired parametrizations of the `pomeron' content,  have been proposed \cite{STIR,COLL,ELL}. More recently a formidable set of data has been presented by the H1 collaboration, and as in the leading-proton case, the comparison between model predictions and data has been found to be rather poor unless a large number of additional elements is included \cite{H1D}.

In the language of fracture functions, both the leading-proton and the diffractive regimes are complementary features of a more general semi-inclusive process. The approach, then, suggests and provides a bridge between the two regimes, which is particularly appropriate and even necessary, at least in the kinematical region where neither the `pomeron' nor the `reggeon' exchange picture alone are expected to describe the data, say $x_{\pp} \sim 0.1$.\\

\noindent{\large \bf Parametrization}\\

Having introduced the arguments for a combined analysis of both leading-proton and diffractive cross sections using the framework of fracture functions, we proceed in this section to their phenomenological determination by means of a LO QCD global analysis of the available data\footnote{Since the data were obtained under the assumption of a negligible longitudinal diffractive structure function, we keep the analysis at leading order for consistency.}.

In order to obtain a parametrization for the proton-to-proton fracture function
$M_2^{p/p}(\beta,Q^2_0,x_{\pp})$ at a given initial scale $Q^2_0$, we  select in the first place, a relatively simple functional dependence in the variables $\beta$ and $x_{\pp}$, but with enough flexibility as to reproduce the data accurately. If one were only interested in the diffractive regime,
the natural choice would be a simple `pomeronic' flux in $x_{\pp}$ times an ordinary parton distribution in $\beta$, namely
\begin{equation}
M_2^{p/p}(\beta,Q^2_0,x_{\pp}) \sim  C_{\pp} x_{\pp}^{\alpha_{\pp}} \times  \,N_{s} \,\beta^{a_s}(1-\beta)^{b_s} \, ,  
\end{equation}
where the label $s$  stands for singlet contribution, assumed to be the dominant one. Analogously, for the leading-proton regime the
natural choice would be almost the same, but with a standard meson or `reggeon' flux or even better, something combining their effects \cite{SZC}:  
\begin{equation}
M_2^{p/p}(\beta,Q^2_0,x_{\pp}) \sim  C_{LP} \, x_{\pp}^{\alpha_{LP}}(1-x_{\pp})^{\beta_{LP}} \times \, \,N_{s} \,\beta^{a_s}(1-\beta)^{b_s} \, .  
\end{equation}

These kinds of parametrizations give relatively good initial approximations to the description of the corresponding data sets; however their survival  seems unlikely  in a more precise analysis. For example the `pomeron' flux factorization hypothesis (independence with respect to $\beta$ and $Q^2$) is a rather strong assumption, not even true for the kinematical range of the available  data. This has been been shown in recent analyses \cite{H1D} and in fact is apparent from fig. 1 where there is a clear $x_{\pp}^{\alpha}$ behaviour for large values of $\beta$, with $\alpha \sim -0.25$, behaviour that is then  suppressed for intermediate values,
and finally is changed to something like a positive power of $x_{\pp}$ at low $\beta$.
       
In order to take into account small departures from the initial approximations, and also combine the two behaviours in such a way that for  low $x_{\pp}$ (diffractive regime)  the `pomeron' picture dominates, while for low  $\beta$ and large $x_{\pp}$ the meson or `reggeon' exchange picture emerges, we propose a modified flux such that the light-quark singlet component $(M_q^{p/p} \equiv 3 M_u^{p/p}= 3 M_d^{p/p}=3 M_s^{p/p})$ of the fracture 
function is parametrized as
\begin{eqnarray}
x M_q^{p/p}(\beta,Q^2_0,x_{\pp}) =   \,N_{s}\, \beta^{a_s}(1-\beta)^{b_s}  \times \,\,\,\,\,\,\,\,\,\,\,\,\,\,\,\,\,\,\,\,\,\,\,\,\,\,\,\,\,\,\,\,\,\,\,\,\,\,\,\,\,\,\,\,\,\,\,\,\,\,\,\,\,\,\,\,\,\,\,\,\,\,\,\,\,\,\,\,\,\,\,\,\,\,\,\,\,\,  \nonumber \\
 \left\{ C_{\pp}\, \beta \, x_{\pp}^{\alpha_{\pp}} + 
C_{LP} \, (1-\beta)^{\gamma_{LP}} \, (1+a_{LP}(1-x_{\pp})^{\beta_{LP}}) \right\} ,
  \end{eqnarray} 
and similarly for gluons with the corresponding parameters $N_g$, $a_g$ and $b_g$.

Notice that  this is only done for the sake of convenience; even though at the initial scale $Q^2_0$ the parametrization implies some sort of flux factorization, beyond the initial scale, the evolution equations drive the fracture function as a whole making the usual discrimination between `fluxes' and `parton densities of the exchanged object'  somewhat ambiguous. This is particularly relevant in this case, where none of the `pomeron' and the `reggeon' contributions are known,  making it then possible  to shift part of
 the pomeron-like contribution into the other during the fitting procedure by simply adjusting the parameters conveniently.

Even at the initial scale, the flux ambiguity is reflected in the fact, that for parametrization purposes, the three normalizations, $N_s$, $C_{\pp}$, and $C_{LP}$, are not independent,  and consequently, any of them can be  
set to unity (we take $C_{\pp}=1$). We do not assume any  momentum sum rule for the distributions in this analysis, since its significance is not clear  even when considering the process in terms of the exchange of `pomerons' and `reggeons'.
 
Regarding the evolution, we set $Q_0^2=2.5\,$GeV$^2$, take $\Lambda_{QCD}=0.232\,$GeV$^2$, and we choose to work in a scheme with a variable number of flavours, where charm and bottom distributions are radiatively generated from their corresponding thresholds (taken as their masses). The main effect of the heavy quarks in the fit is to reduce the light-quark singlet  component, effect that is particularly sizeable for charm quarks due to the smaller mass (almost coincident with the initial scale), and  the influence of the large gluon distribution in the evolution.

In the analysis of both H1 data sets \cite{H1D,H1LP}, which were obtained within different ranges of the variable $t$,
we have assumed the universal validity of the exponential behaviour in $t$ measured by the ZEUS collaboration \cite{ZD} and have rescaled the data to a common range, given by that of the diffractive data of H1 ($-M_p^2 x_{\pp}^2/(1-x_\pp) \equiv-t_{min}<-t<1$ GeV$^2$). 
 The parametrization obtained through the fitting procedure corresponds to that particular range of the variable $t$; it has to be rescaled for the computation of any other observable if  different kinematical cuts are applied. 
This is particularly important for the leading-proton H1 data, since a small transverse momentum cut    was implemented $p_T < 200$ MeV, selecting only data corresponding to $|t|\ll 1$ GeV$^2$. Such a small cut leads to a  large rescaling factor (close to 3). We have also taken into account an additional error of   5\% for the leading-proton data, because the uncertainties in the rescaling process. 

Phenomenological parametrizations like the one in eq. (10) lead to simultaneous QCD fits for both the diffractive and the leading-proton cross sections obtained by H1 with very good precision.
In order to ensure that the evolution code is working properly in the region of large $\beta$, we imposed a lower constraint on the $(1-\beta)$ exponents for both quark and gluon distributions,  $b_s, \, b_g > 0.1$.
Furthermore, in the expectation of very hard distributions we saturate this constraint and fix $a_s=0$ (leaving only 8 free parameters). Doing this we obtain a global fit with $\chi_{total}^2/$d.o.f.=1.09 ($\chi^2=292.23$, data points $=274$),   which we designate by Fit A. 

Similar parametrizations, but with softer gluons, yield slightly higher values, for example in Fit B, $b_g$ is set to $0.7$ (whereas $b_s$ remains 0.1)  finding $\chi_{total}^2/$d.o.f.=1.13. 
The values found for the parameters of both sets at the initial scale $Q_0^2$ are given in table 1.

\begin{center}
\begin{tabular}{|c|c|c|c|c|c|c|c|c|} \hline \hline
Set&$\alpha_{\pp}$&$C_{LP}$&$\beta_{LP}$&$\gamma_{LP}$&$a_{LP}$&$N_s$ &$N_g$ &$a_g$ \\ \hline
A & $-$1.260 & 14.395 & 32.901&2.627&12.320& 0.041& 0.354 & 0.450 \\  \hline
B & $-$1.257 & 12.556 & 32.412&2.338&11.412& 0.047& 0.694 & 0.648 \\  \hline \hline
\end{tabular} 
\vspace*{2mm} 
\end{center} 
\begin{center} {\footnotesize {\bf TABLE 1:}  Parameters for $Q_0^2=2.5\, $GeV$^2$.} 
\end{center}

As   is shown by the solid lines in figs. 1 and 2, the accuracy of the fit is remarkably good in the case of H1 diffractive data ($\chi_{H1D}^2/data=215.63/226$), and also good for the leading-proton data. The dashed line in fig. 1 shows the contribution coming from the $x_{\pp}^{\alpha_{\pp}}$ term in eq. (10), which could be interpreted as the `pomeronic component' of the fracture function. This contribution is clearly dominant for large $\beta$, but fails to do so in the low-$\beta$ and large-$x_{\pp}$ region where the contribution from the second term in eq. (10) changes the $x_{\pp}$ slope of the distributions. 

In fig. 3, we show the light-quark singlet and gluon  fracture densities at the initial scale $Q_0^2$  and at an intermediate value of $Q^2=10\,$GeV$^2$ for two characteristic values of $x_{\pp}$. The first one, $x_{\pp}=0.005$, which corresponds to well inside the diffractive regime, shows a rather hard behaviour, whereas the second one, $x_{\pp}=0.2$, which belongs to the 
leading-proton regime, is much softer. It is worth noticing that, at large $x_{\pp}$, the large-$\beta$ behaviour of the distributions is not well constrained because  leading-proton data are only available in the very small-$\beta$ range.

As can be seen from the plots, the differences in the initial distributions between fits A and B tend to be washed out by the evolution and are mostly sizeable for the gluon  at $\beta>0.5$. 
As also shown in fig. 3, gluons carry much more impulse than quarks, specially in the case of the diffractive regime and at low scales. The evolution damps down this gluon dominance, suggesting a large fraction of valence-like gluons
in the `pomeron', which is not so apparent for `reggeons'.  
Putting both mechanisms together in the physical fracture function (fig. 4), the ratio between the `momenta' carried by gluons and quarks\footnote{This ratio is defined as the one between  the integrals of $x\, M_g$ and $x\, M_q$ in the range of $\beta$ $[0,1]$. In this way, when either of the exchange pictures dominates,  the ratio coincides with the usual definition for the fraction of the momentum carried by partons in either  the `pomeron' or the `reggeon'.}
 depends strongly on the value of $x_{\pp}$, which makes   one or the other dominant.
Even though gluons clearly dominate, parametrizations with only gluons at the initial scale, or even at lower values ($Q^2_0=1\,$GeV$^2$), and thus having only quarks of radiative origin, lead to global fits of considerably poorer quality ($\chi^2/$d.o.f. $> 1.5$)  

The main conclusion that can be drawn about the gluon density from the   analysis is that   the
 distribution is important at large $\beta$, at variance with the one for inclusive structure functions, but  its behaviour cannot be precisely determined yet from the available data,  in particular, the exponent  of the $(1-\beta)$ factor in the parametrization.\\

\noindent{\large \bf Scale dependence}\\

As it has been said, the fracture function approach leads to very definite predictions about the scale dependence of the  semi-inclusive cross sections. The rigorous factorization of the cross sections achieved within
this formalism allows a precise perturbative QCD analysis of the scale dependence of the data, as is usually done for ordinary structure functions. 

In figs. 5 and 6 we compare H1 diffractive and leading-proton data, at fixed values of $x_{\pp}$ and $\beta$, as a function of $Q^2$, with the evolved fracture function. The scale dependence induced by the evolution equations (5) is perfectly consistent with the data. For very large values of $\beta$ ($\beta \sim 0.9$) there seems to be some evidence of higher-twist contributions, as has been predicted within some perturbative models \cite{GEN}. 

It is also worth noticing that within the measured range, the scale dependence is dominated by the homogeneous evolution. The effects of the inhomogeneous term become important only for low values of $x_L$ where fragmentation functions are larger. These effects are, however, beyond the range of present data, as can be seen in fig. 7. There, the solid lines represent the scale dependence induced in the fracture function by ordinary evolution equations (homogeneous term in eq. (5)), whereas the dashed lines take into account both contributions.
In order to compute the inhomogeneous contribution we have used GRV94 parton distributions \cite{GRV94}, and the  fragmentation functions of ref. \cite{BIN}.   \\

\noindent{\large \bf ZEUS measurements}\\

In order to check the distributions obtained in the previous section, and also
our assumption about the $t$-dependence, in the following we compare results obtained using our best parametrization (Fit A) with data sets presented by the ZEUS collaboration, which had not been included in the fit.

First, we compare ZEUS diffractive measurements \cite{ZD} with the outcome
of the parametrization after the appropriate evolution to the mean scale value of the data $Q^2=8 \,$GeV$^2$, and the already mentioned rescaling in $t$. Figure 8 shows the agreement found between the data and the fracture function estimate for different bins in $\beta$. 

It is important to stress that at variance with H1 diffractive data, the ZEUS data are obtained by identifying the final-state proton, i.e.  explicitly  collecting semi-inclusive events. The agreement found gives us confidence in the assumption about the dominance of single dissociative processes $\gamma^*p \rightarrow Xp$ in H1 diffractive data, and then the applicability of a semi-inclusive framework.

Going to leading-proton data, the ZEUS collaboration has measured the fraction of DIS events with a leading-proton in the final state giving special attention to the dependence in the variable $t$ \cite{ZLP}. At least two different sets of data have been presented, corresponding  to different ranges in $t$. After the adequate rescaling   of the parametrization for the fracture function at an average value of $Q^2=10$ GeV$^2$, and using GRV94 \cite{GRV94} parton distributions for the estimate of the total number of events \cite{NEUTR}, we obtain a remarkable agreement with the data in the common $x_L$-range, as shown in figs.  9a and 9b. The $Q^2$ dependence is found to almost cancel in the ratio between the corresponding fracture and structure functions.
 
Again, the general framework, and particularly the assumption about the $t$ dependence, seems to be nicely confirmed. Notice also  the fact that the parametrization interpolates fairly well  the diffractive and leading-proton behaviour between H1 kinematical regions (thick lines) where the data used in the fit belong. 

The resulting interpolation is radically different from those obtained, for example, adjusting diffractive data with a purely `pomeronic' flux, indicating the relevance of other contributions in the region $x_{\pp}\sim 0.05$--$0.1$. These effects are of extraordinary importance when addressing the comparison between 
diffractive DIS and $p\overline{p}$ collision data as a mean of testing factorization \cite{COLL}. \\

\noindent{\large \bf Conclusions}\\

We have shown that an approach based on fracture functions motivates, from a formal point of view, and also phenomenologically  allows a unified
description of both leading-proton and diffractive deep inelastic cross sections. A simple parametric form for this function gives a very accurate
description of the  data  available at present providing a smooth interpolation between the distinctive behaviours of the two regimes, also in accordance with ZEUS data. The analysis also hints at  some non-perturbative features,  such as a strong-gluon dominance in the `pomeronic' component with a characteristic valence-like behaviour but with still large uncertainties in the large-$\beta$ region. Finally, our results  verify that the scale dependence of the data agrees with the one
predicted by the fracture function formalism.

A code containing  both sets of parametrizations can be requested from the authors by e-mail at  daniel.deflorian@cern.ch or sassot@df.uba.ar\\

\noindent{\large \bf Acknowledgements}

We warmly acknowledge C. A. Garc\'\i a Canal for interesting discussions, N. Cartiglia for helpful comments and suggestions, and F. Barreiro for his hospitality during the Madrid Workshop on Low-x Physics, where this work began.  The work of D.deF. was partially supported by the World Laboratory (project T1) and by the EU Fourth Framework Programme `Training and Mobility of Researchers', Network `Quantum Chromodynamics and the Deep Structure of Elementary Particles', contract FMRX-CT98-0194 (DG 12-MIHT). 

\pagebreak

\pagebreak

\setlength{\unitlength}{1.mm}
\begin{figure}[hbt]
\begin{picture}(170,150)(0,0)
\put(-5,-35){\mbox{\epsfxsize16.0cm\epsffile{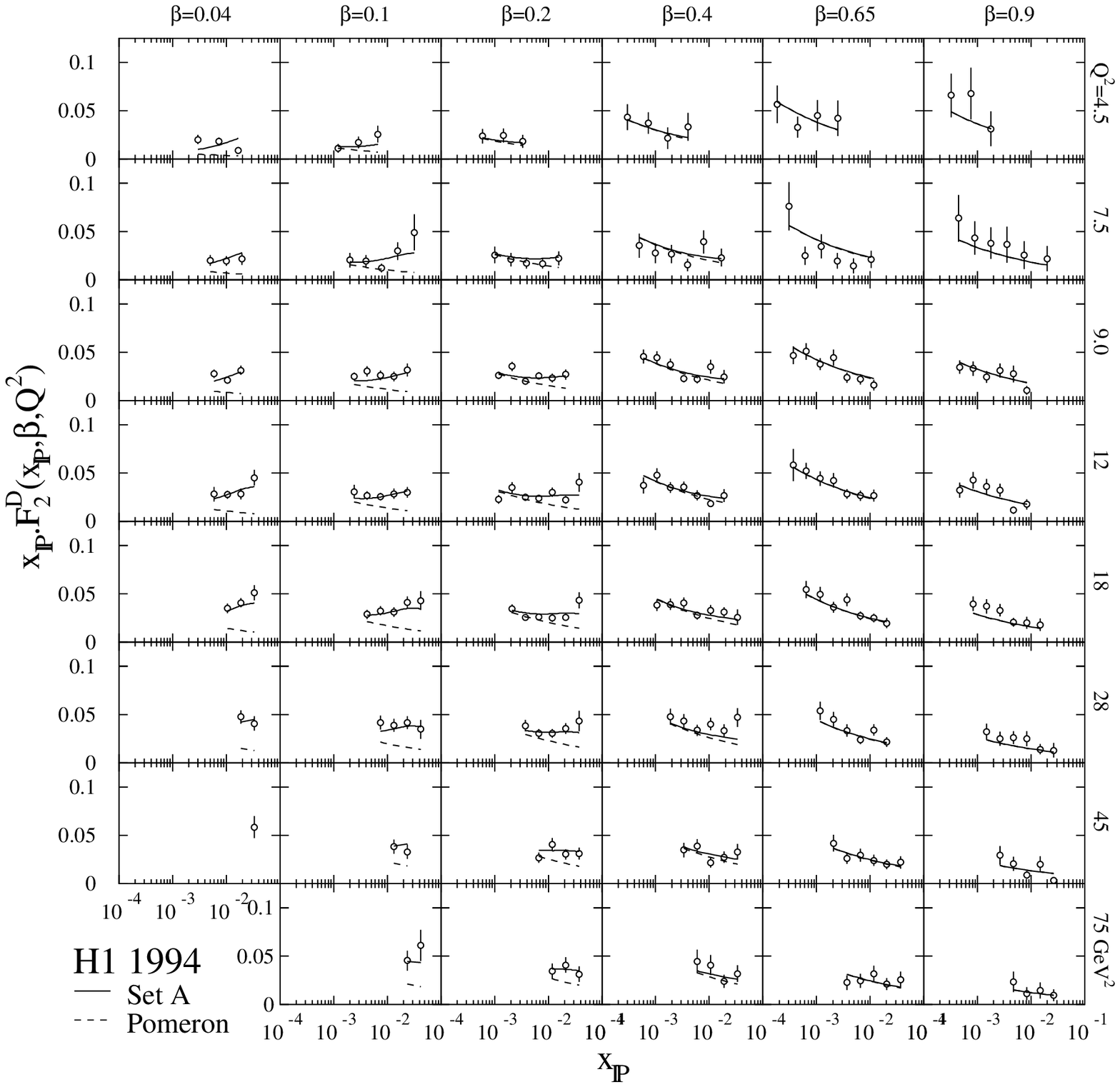}}}
\put(0,-15){\mbox{\bf Figure 1:}{  H1 diffractive data against the outcome of the fracture function}}
\put(0,-20){\mbox{ parametrization (solid lines) and its pomeron-like component (dashed lines).
 }}
\end{picture}
\end{figure}

\setlength{\unitlength}{1.mm}
\begin{figure}[hbt]
\begin{picture}(160,150)(0,0)
\put(-5,-35){\mbox{\epsfxsize16.0cm\epsffile{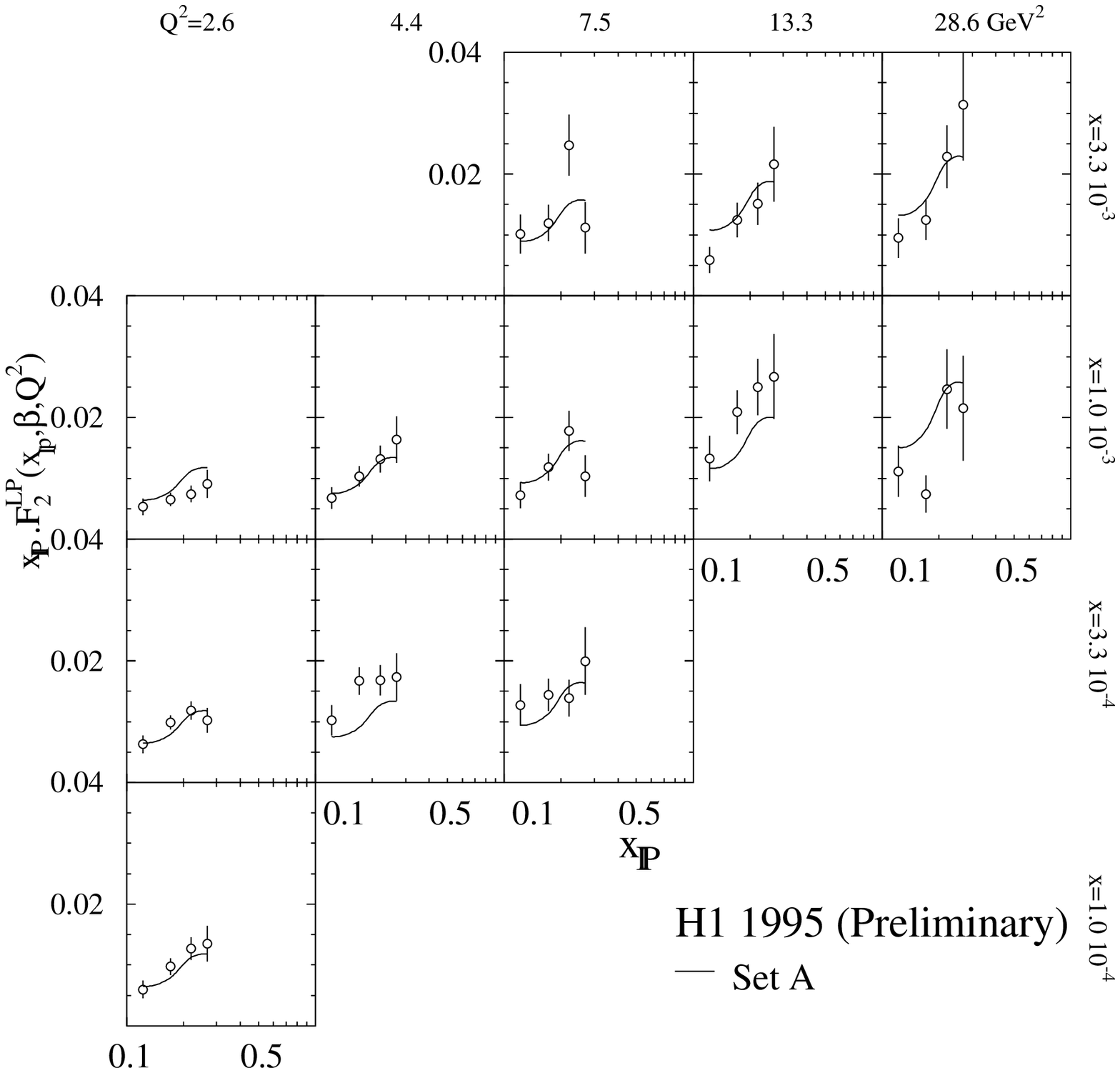}}}
\put(0,-15){\mbox{\bf Figure 2:}{ H1 leading-proton data against the outcome of the fracture function}}
\put(0,-20){\mbox{ parametrization (solid lines).}}
\end{picture}
\end{figure}

\pagebreak
 
\setlength{\unitlength}{1.mm}
\begin{figure}[hbt]
\begin{picture}(170,150)(0,0)
\put(-5,-35){\mbox{\epsfxsize15.0cm\epsffile{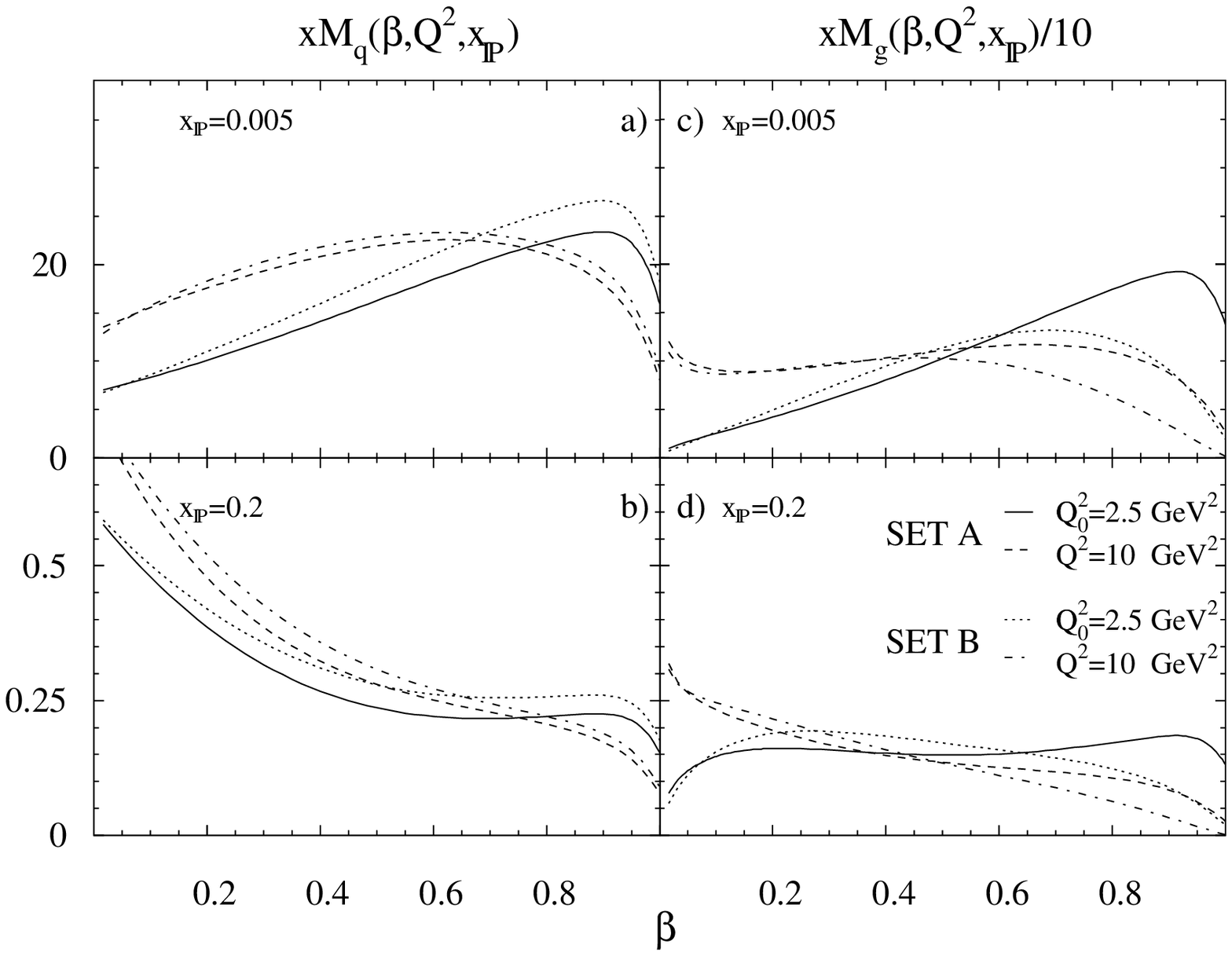}}}
\put(0,-15){\mbox{\bf Figure 3:}{ Fracture function densities: a) and  b) light-quark singlet, }}
\put(0,-20){\mbox{c) and d) gluons. }}
\end{picture}
 \end{figure}

\setlength{\unitlength}{1.0mm}
\begin{figure}[hbt]
\begin{picture}(170,150)(0,0)
\put(-5,-25){\mbox{\epsfxsize16.0cm\epsffile{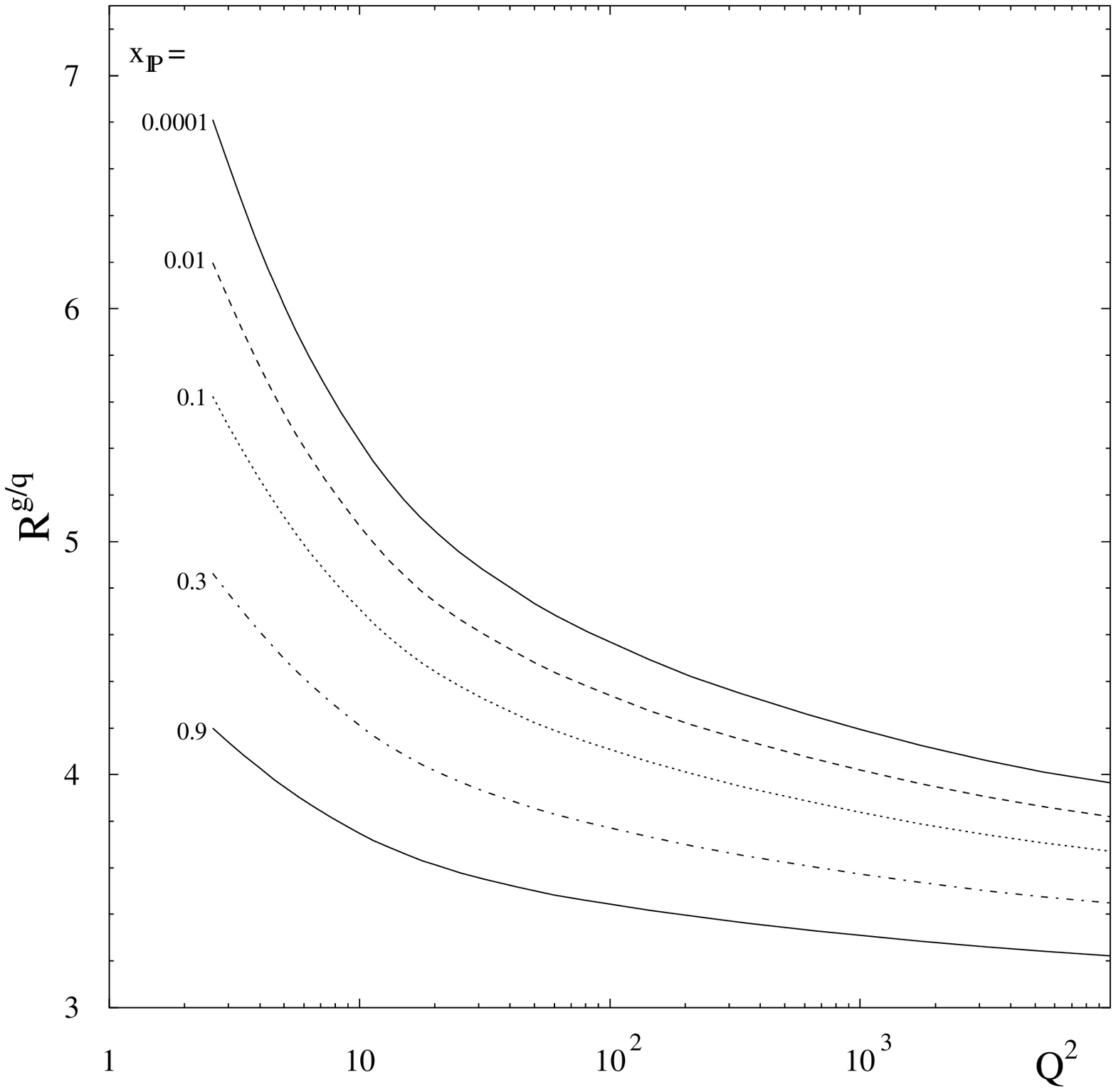}}}
\put(-5,-15){\mbox{\bf Figure 4:}{ Ratio between the total momentum carried by gluons and quarks (Fit A).}}
 \end{picture}
\end{figure}

\setlength{\unitlength}{1.mm}
\begin{figure}[hbt]
\begin{picture}(170,150)(0,0)
\put(-5,-35){\mbox{\epsfxsize16.0cm\epsffile{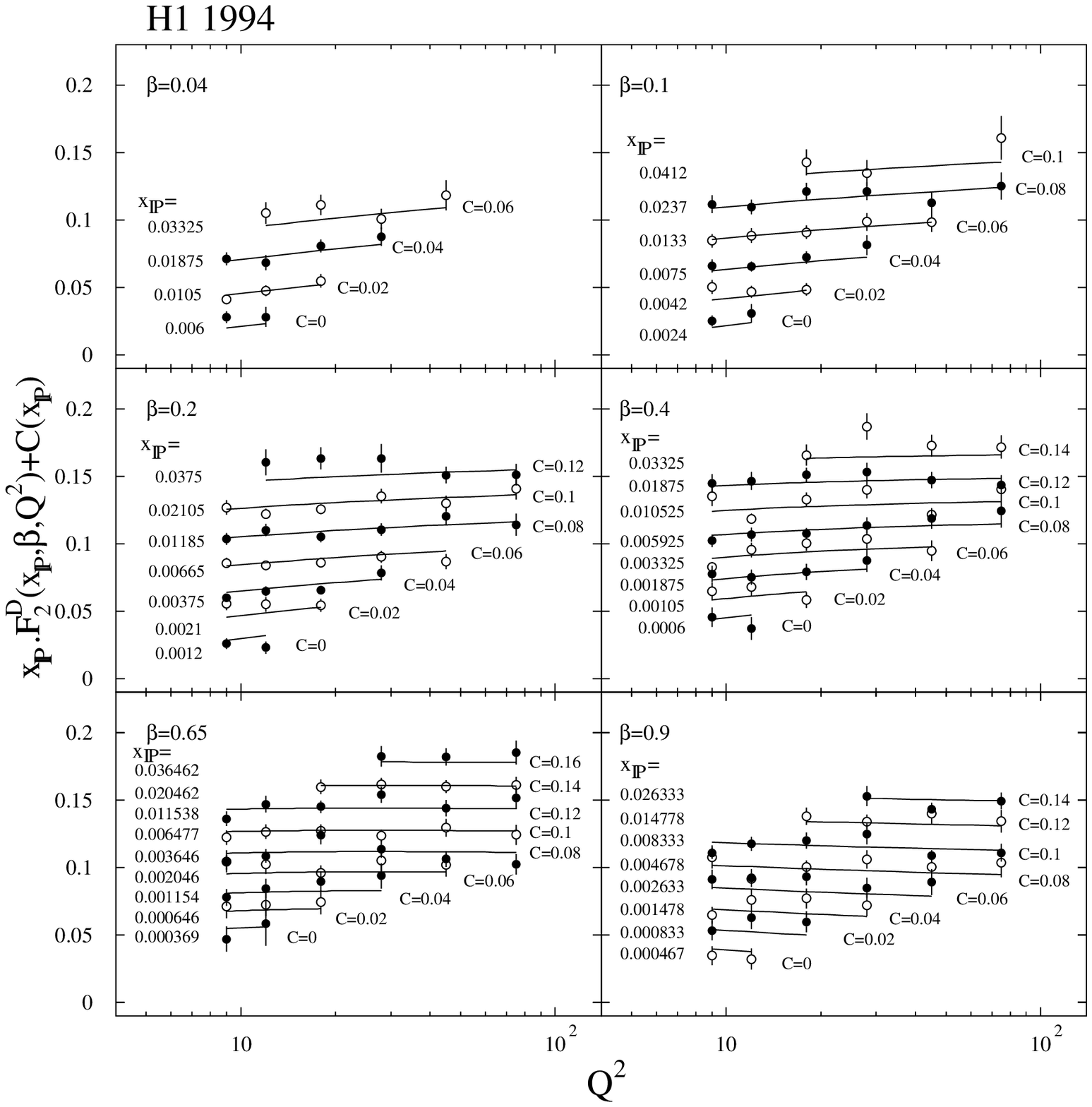}}}
\put(-0,-15){\mbox{\bf Figure 5:}{ Scale dependence of H1 diffractive data and the one obtained}}
\put(-0,-20){\mbox{evolving the Fit A parametrization (Fit B renders undistinguishable results).
}}
\end{picture}
\end{figure}

\setlength{\unitlength}{1.mm}
\begin{figure}[hbt]
\begin{picture}(170,150)(0,0)
\put(-5,-35){\mbox{\epsfxsize16.0cm\epsffile{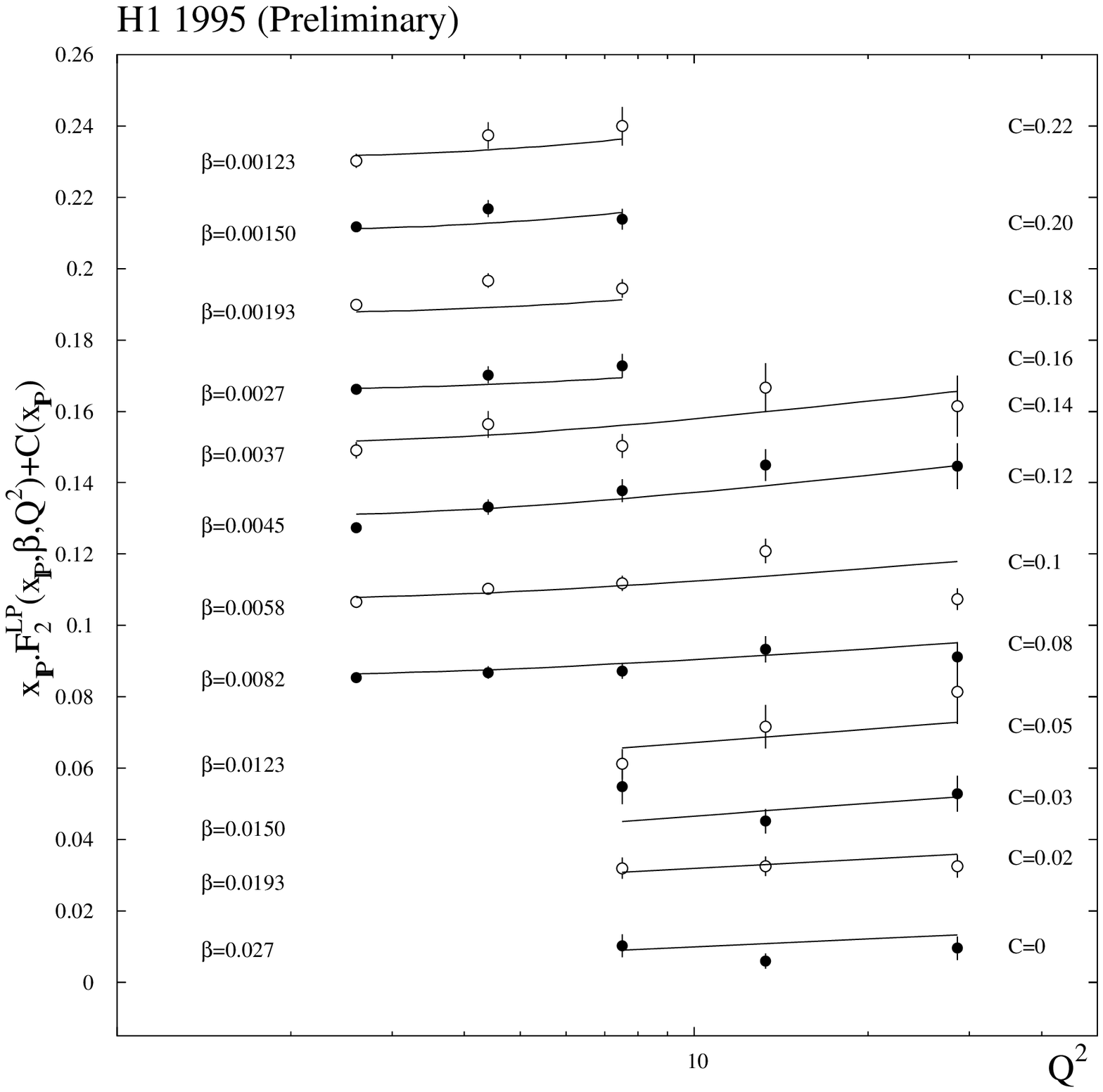}}}
\put(0,-15){\mbox{\bf Figure 6:}{ Scale dependence of H1 leading-proton data and the one obtained}}
\put(-0,-20){\mbox{evolving the Fit A parametrization (Fit B renders undistinguishable results).}}
\end{picture}
\end{figure}

\setlength{\unitlength}{1.mm}
\begin{figure}[hbt]
\begin{picture}(170,150)(0,0)
\put(-10,-35){\mbox{\epsfxsize16.0cm\epsffile{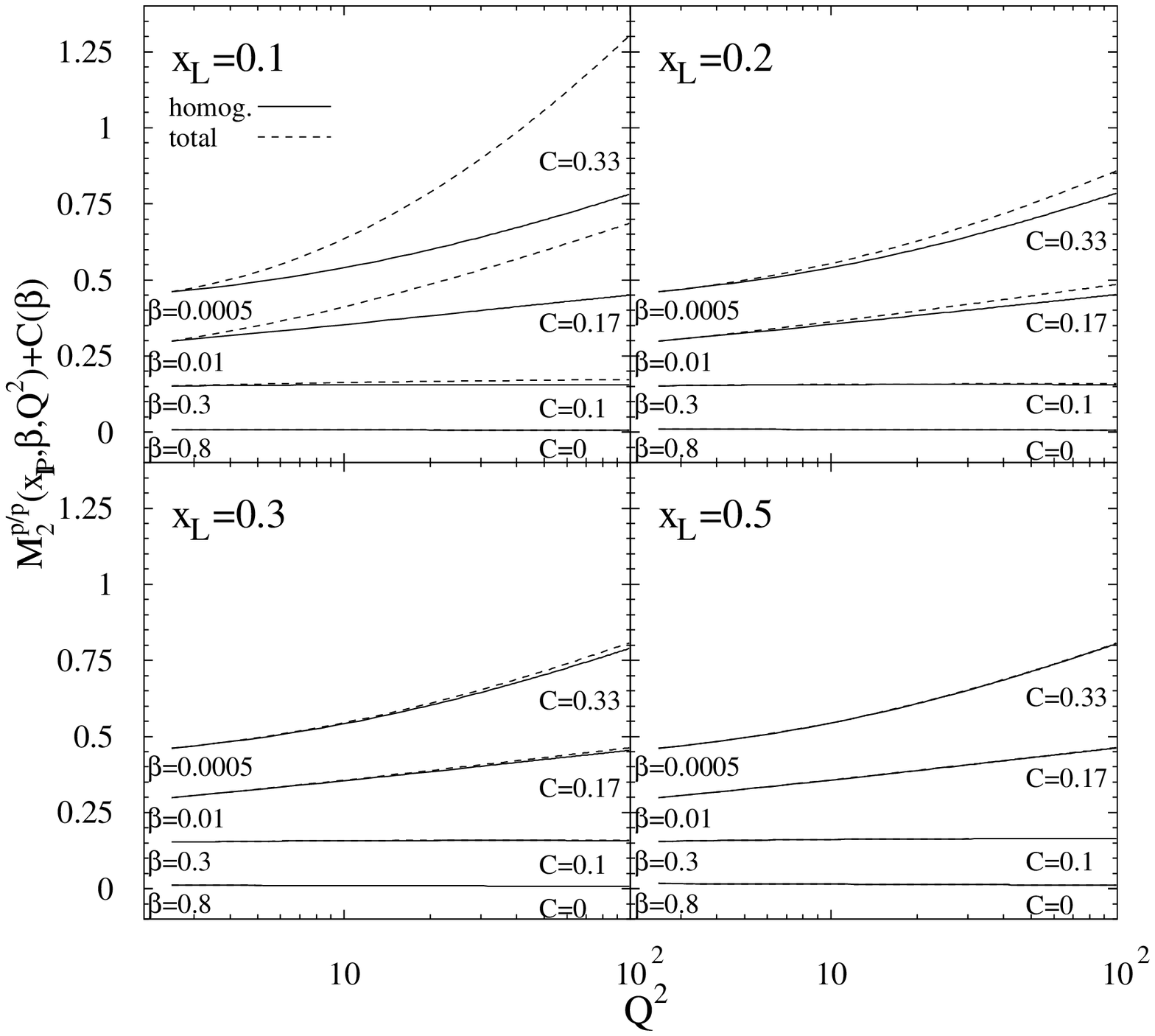}}}
\put(0,-10){\mbox{\bf Figure 7:}{ Homogeneous and complete evolution of the proton-to-proton}}
\put(-0,-15){\mbox{fracture function (solid and dashed lines, respectively).}}
\end{picture}
\end{figure}

\setlength{\unitlength}{1.mm}
\begin{figure}[hbt]
\begin{picture}(170,150)(0,0)
\put(-10,-35){\mbox{\epsfxsize16.0cm\epsffile{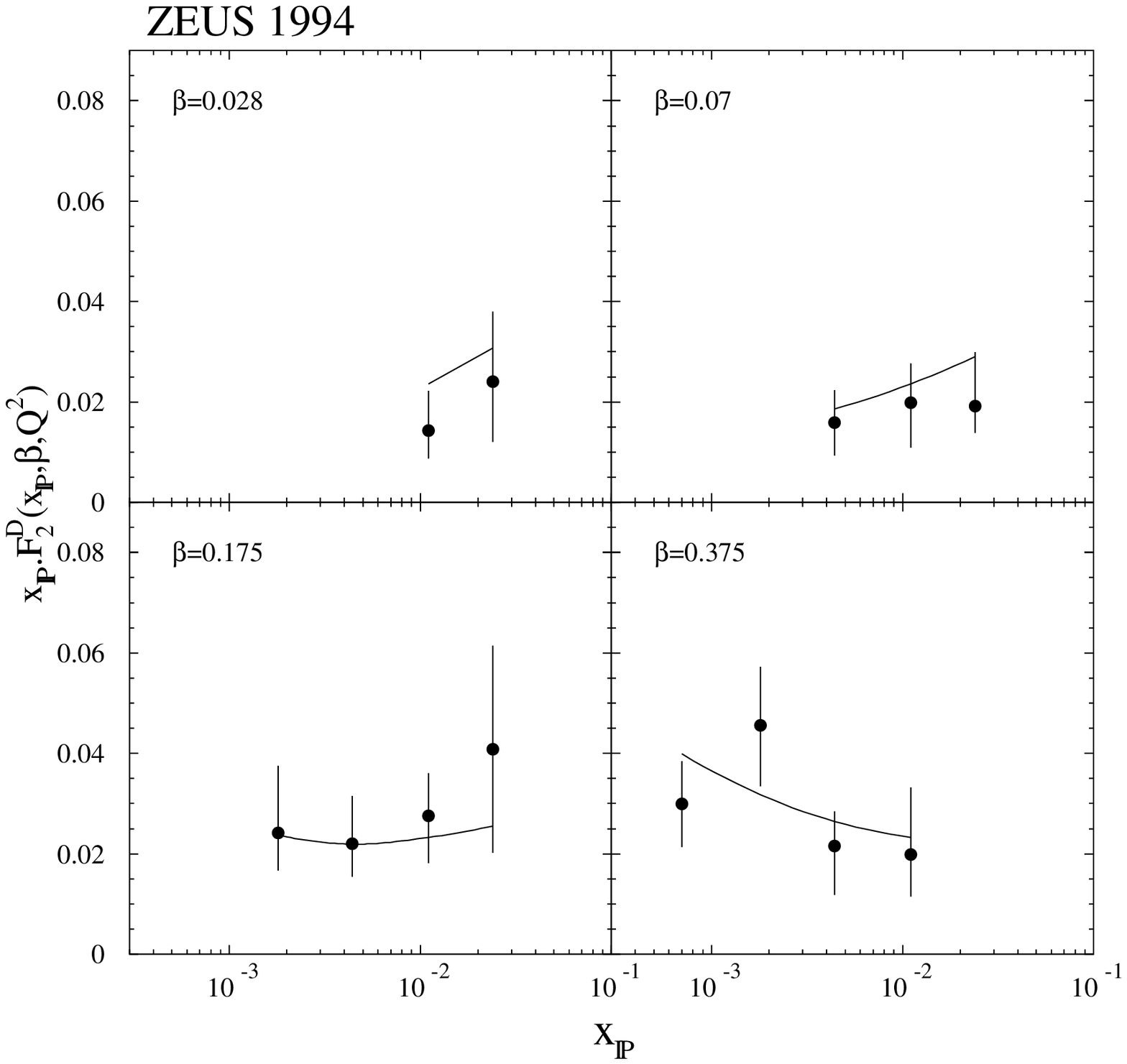}}}
\put(0,-5){\mbox{\bf Figure 8:}{ ZEUS diffractive data, against the expectation coming from the }}
\put(-0,-10){\mbox{fracture function parametrization (Fit A).}}
\end{picture}
\end{figure}

\setlength{\unitlength}{1.0mm}
\begin{figure}[hbt]
\begin{picture}(170,150)(0,0)
\put(-5,-20){\mbox{\epsfxsize15.0cm\epsffile{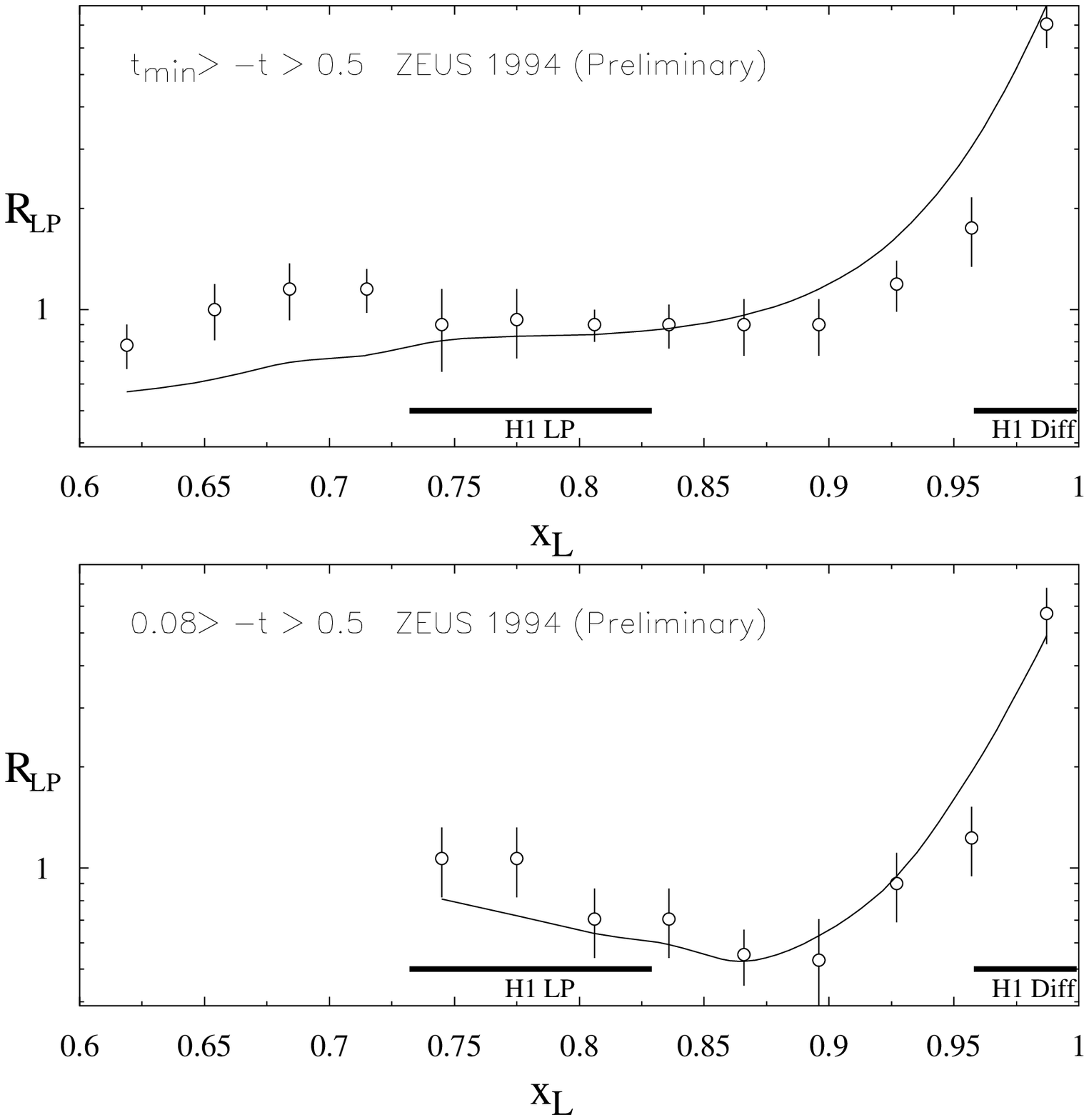}}}
\put(0,-10){\mbox{\bf Figure 9:}{ The fraction of DIS events with a 
leading-proton for different}}
\put(-0,-15){\mbox{ranges of $t$ as measured by ZEUS, against the expectation coming from }}
\put(-0,-20){\mbox{the fracture function parametrization (Fit A).}}
\end{picture}
\end{figure}

\end{document}